\documentclass[twocolumn,prb,aps]{revtex4}
\usepackage[dvips]{graphicx}
\usepackage{longtable}
\begin{document}
\title{Analytical continuation of imaginary axis data using maximum entropy}

\pacs{}

\begin{abstract} 
We study the maximum entropy (MaxEnt) approach for analytical continuation 
of spectral data from imaginary times to real frequencies. The total error is
divided in a statistical error, due to the noise in the input data, and a 
systematic error, due to deviations of the default function, used in the 
MaxEnt approach, from the exact spectrum. We find that the MaxEnt approach
in its classical formulation can lead to a nonoptimal balance between
the two types of errors, leading to an unnecessary large statistical error. The 
statistical error can be reduced by splitting up the data in several batches,
performing a MaxEnt calculation for each batch and averaging. This can   
outweigh  an increase in the systematic error resulting from this approach. 
The output from the MaxEnt result can be used as a default function for 
a new MaxEnt calculation. Such iterations often lead to worse results
due to an increase in the statistical error. By splitting up the data
in batches, the statistical error is reduced and and the increase resulting
from iterations can be outweighed by a decrease in the systematic error.
Finally we consider a linearized version to obtain a better understanding 
of the method.

\end{abstract}

\author{O. Gunnarsson$^{(1)}$, M. W.  Haverkort$^{(1)}$ and G. Sangiovanni$^{(2)}$}   
\affiliation{
${}^1$Max-Planck-Institut f\"ur Festk\"orperforschung, D-70506 Stuttgart, Germany  \\
${}^2$Institut f\"ur Festk\"orperphysik, Technische Universit\"at Wien, Vienna, Austria
}

\maketitle

\section{Introduction}\label{sec:1}

The analytical continuation of spectral functions from imaginary time $\tau$
to real energies $\omega$ is a difficult problem
due to its ill-posed nature, i.e., the output can depend very sensitively on the input. For strongly correlated electrons,
however, this is an important problem.  Most approaches for such systems 
involve uncontrolled approximations. Using quantum Monte-Carlo (QMC) methods 
or quantum cluster methods it is, however, possible to obtain accurate data 
for Green's functions and response functions on the imaginary axis, raising
the problem of analytical continuation to the real axis. Since these methods 
provide data with statistical noise, the ill-posed nature of the problem makes 
analytical continuation very difficult. 

This problem can be treated within the Bayesian 
theory.\cite{Jarrell,MEMref} The problem is regularized by introducing 
an entropy in terms of the deviation of the output real axis spectrum 
from some default function. The importance of the entropy is controlled 
by a parameter $\alpha$, which is determined using statistical 
arguments.\cite{Jarrell,MEMref} This method is referred to as the Maximum
Entropy (MaxEnt) method. It has been rather successful in performing analytical
continuations. Alternative methods have been proposed, such as Pad\'{e}
approximations,\cite{Vidberg,Baker} singular value decomposition,\cite{svd} 
stochastic regularization\cite{Rubtsov} and sampling schemes.\cite{Mishchenko,Kiamars} 

In this paper we focus on the MaxEnt method. This method is usually discussed 
in terms of the Bayesian theory. Here we start from the equations 
generated by the MaxEnt formalism and use an algebraic approach to analyze 
the theory. We discuss the accuracy that can be obtained within this framework. 
The error in the output spectral function can be split up in a statistical 
error, due to the noise in the input data, and a systematic error, due
to the deviation of the default function from the true spectrum. The
choice of $\alpha$ determines the relative size of these errors. In the 
classical MaxEnt method the most probable $\alpha$ is chosen.\cite{Jarrell}
We find that this choice can make the statistical error unnecessary large.

The input data is typically given as a number $N_{\rm sample}$ of samples, 
$\bar G_{\nu}(\tau)$, where each sample gives a (noisy) version of the 
imaginary time function $G(\tau)$.  
We find that the accuracy can sometimes be improved by splitting 
up the samples in $N_{\rm calc}$ sub sets (batches), with $N_{\rm sample}/N_{\rm calc}$
samples in each batch.  We then perform $N_{\rm calc}$ MaxEnt calculations, each 
with $N_{\rm sample}/N_{\rm calc}$ samples, and then average the results, instead 
of performing one MaxEnt calculation $N_{\rm sample}$ samples. This approach 
reduces the statistical error at the cost of an increase in the systematic error. 

We also discuss the possibility of an iterative MaxEnt method, where the 
output is used to define a new default function. This usually works poorly, 
and we show that this is due to an increase in the statistical error, 
overwhelming the improvement in the systematic error.  However, if the data 
are split in batches,  as discussed above, the importance of the statistical 
error can be reduced to the point where the approach improves the total accuracy. 

To further analyze the results, we introduce an alternative method with a 
new, slightly different definition of the entropy. This leads to a set of 
linear equations, where the propagation of the errors can be analyzed 
more easily and features of the MaxEnt method better understood. This method, 
however, does not guarantee a positive spectral function, and it is less 
useful for practical calculations.

\begin{figure}
{\rotatebox{-90}{\resizebox{6.cm}{!}{\includegraphics {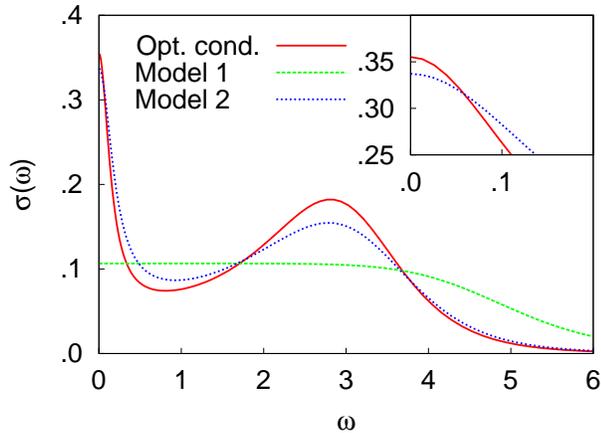}}}}
\caption{\label{fig:1}(color on-line) ``Exact'' spectral function and two
different default models as a function of frequency. The inset shows the
models on a small energy scale. 
}
\end{figure}
In this paper we focus on a response function, the optical conductivity 
$\sigma(\omega)$. We introduce a typical $\sigma(\omega)$, which in the
following will be refered to as the ``exact'' $\sigma(\omega)$. The form of
$\sigma(\omega)$ was chosen using results for the two-dimensional Hubbard model as a guide. 
This model of $\sigma(\omega)$ 
can easily and accurately be transformed to imaginary axis data. We add 
statistical noise to the data and then transform the data back to the 
real axis, using the various modifications of the MaxEnt method. If a 
given method worked perfectly the $\sigma(\omega)$ that we started with
should be recovered exactly. The deviations from the ``exact'' 
$\sigma(\omega)$ are then a measure of the 
accuracy of the different approaches. 

As an example, Fig.~\ref{fig:1}
shows an ``exact'' optical conductivity and two default models used 
in the MaxEnt approach. The optical conductivity has a Drude like peak at 
$\omega=0$ and a ``Hubbard'' peak at $\omega \sim 3$ corresponding to transitions
between the Hubbard bands. The default models are chosen so that they 
satisfy the exact sum rule. The two models are chosen according to 
two different strategies. It is sometimes argued that the default model should 
contain little information, apart from  certain exact results, such as sum
rules. This way the results are not prejudiced by possible incorrect 
assumptions. Model 1 has been chosen this way. Alternatively, as a default
model one can use the output from a calculation at a higher temperature $T$.
Model 2 has therefore been chosen to be quite similar to the ``exact''     
result, but with all features somewhat broader. Model 2 will naturally 
deliver much more accurate output spectra.                        

The paper is organized as follows.
In Sec.~\ref{sec:2} we introduce the formalism. Sec.~\ref{sec:3} describes 
how a MaxEnt calculation is performed as an average of several MaxEnt
calculations and Sec.~\ref{sec:4} discusses an iterative MaxEnt method.
In Sec.~\ref{sec:5} we present a simplified entropy definition, leading 
to linear equations.

\section{Formalism}\label{sec:2}

We introduce the basic formalism, essentially following Jarrell and 
Gubernatis,\cite{Jarrell} and then provide error estimates.
The function $G_i=G(\tau_i)$ for imaginary time $\tau_i$ is related 
to a spectral function $A_i=A(\omega_i)$ on the real frequency axis 
$\omega$,   
\begin{equation}\label{eq:a1}
G_i=\sum_{j=1}^{N_{\omega}}K_{ij}A_j \hskip1cm i=1,...N_{\tau},
\end{equation}
via a kernel $K_{ij}=K(\tau_i,\omega_j)$, given for some discrete values
$\omega_j$ of $\omega$. For the case of the optical 
conductivity, considered here, the kernel is given by
\begin{equation}\label{eq:a2}
K_{ij}={1\over \pi}  {\omega_j \over 1-{\rm exp}(-\beta \omega_j)}
(e^{-\omega_j \tau_i}+e^{-(\beta-\tau_i)\omega_j})f_j, 
\end{equation}
where $f_j$ is a weight factor chosen so that Eq.~(\ref{eq:a1}) corresponds
to an integral over $\omega$. For the electron Green's function the 
corresponding kernel is
\begin{equation}\label{eq:a2a}
K_{ij}={e^{-\tau_i\omega_j}\over 1+e^{-\beta \omega_j}}f_j.
\end{equation}
We introduce a likelihood function
\begin{equation}\label{eq:a3}
L={1\over 2}\sum_{i=1}^{N_{\tau}} ({\bar G_i-G_i \over \sigma_i})^2,
\end{equation}
where $\bar G_i$ are data obtained from, e.g., a Monte-Carlo calculation,
with the statistical accuracy  $\sigma_i$, 
and $G_i$ has been calculated from Eq~(\ref{eq:a1}). 
We also introduce the entropy 
\begin{equation}\label{eq:a4}
S=\sum_{i=1}^{N_{\omega}} f_i(A_i-m_i-A_i {\rm ln} {A_i\over m_i}),
\end{equation}
where $m_i$ is a default model. The quantity $L-\alpha S$ is then minimized
with respect $A_j$.  This leads to the equations
\begin{equation}\label{eq:a5}
-\sum_{i=1}^{N_{\tau}}{\bar G_i-G_i\over \sigma_i^2}K_{ij}+
\alpha f_j{\rm ln}{A_j\over m_j}=0
\end{equation}
These equations are solved to obtain the spectral function $A_i$.
The quantity $\alpha$ can be determined using statistical methods,
giving the most probable $\alpha$. This is referred to as the classical
MaxEnt method.\cite{Jarrell} Alternatively, one can average the spectrum
calculated for different values of $\alpha$, using the probability 
of that $\alpha$ as a weighting function.\cite{Jarrell} This
method, Bryan's method, gives similar results for the cases considered 
here.

To estimate the error in this approach, we express the calculated 
spectral function $A$ in terms of the exact result $A^{\rm exact}$ as
\begin{equation}\label{eq:a5a}
A_i=A_i^{\rm exact}+\Delta A_i,
\end{equation}
where $\Delta A_i$ is the error in $A_i$. We assume that the error 
is sufficiently small that the logarithm in Eq.~(\ref{eq:a5}) can
be expanded to lowest order. Then  
\begin{equation}\label{eq:a6}
-\sum_{i=1}^{N_{\tau}}{\Delta \bar G_i-\Delta G_i\over \sigma_i^2}K_{ij}+
\alpha f_j({\rm ln}{A_j^{\rm exact} \over m_j}+
{\Delta A_i \over A_i^{\rm exact}})=0,
\end{equation}
where $\Delta G_i=\sum_j K_{ij}\Delta A_j$ and $\Delta \bar G_i=
\bar G_i-\sum_j K_{ij}A^{\rm exact}_j$ is the error in $\bar G_i$ due to 
the statistical noise. To solve these equations, we define
\begin{equation}\label{eq:a7a}
a_j=\sum_{i=1}^{N_{\tau}}{\Delta \bar G_iK_{ij} \over \sigma_i^2}+\alpha f_j{\rm ln}{m_j \over A^{\rm exact}_j} 
\end{equation}
\begin{equation}\label{eq:a7b}
b_{jk}=\sum_{i=1}^{N_{\tau}}{K_{ij}K_{ik}\over \sigma_i^2}+{\alpha f_j
\over A^{\rm exact}_j}\delta_{jk}.
\end{equation}
Using matrix notations,
\begin{equation}\label{eq:a8}
\Delta A=b^{-1}a=b^{-1}K^{\rm T}\sigma^{-2} \Delta \bar G+b^{-1}\alpha f
{\rm ln} ({m\over A^{\rm exact}}).
\end{equation}
The error $w$ is defined as 
\begin{equation}\label{eq:a8a}
w=\langle \sum_{i=1}^{N_{\omega}}(\Delta A_i)^2 f_i \rangle
\equiv w_{\rm stat}+w_{\rm syst},
\end{equation}
where $\langle  ... \rangle$ denotes the average over many different 
realizations of the noise $\Delta \bar G_i$ in the input data. Here 
$w_{stat}$ is the error due to this noise and $w_{syst}$ is the error 
due to the deviation 
\begin{equation}\label{eq:a8b}
\Delta m_i=m_i-A_i^{\rm exact}
\end{equation}
of the default function from the exact result. Since the noise is random,
there is no contribution from the cross term in the square of the two terms
in Eq.~(\ref{eq:a8}). 
The statistical error can then be written as 
\begin{eqnarray}\label{eq:a14}
w_{\rm stat}=&&\langle \sum_{j,k=1}^{N_{\tau}} {\Delta G_j \over \sigma_j^2}
(Kb^{-1}fb^{-1}K^T)_{jk} {\Delta G_k \over \sigma_k^2}\rangle  \\
=&&\sum_{j=1}^{N_{\tau}} {(Kb^{-1}fb^{-1}K^T)_{jj}\over \sigma_j^2}=
{\rm Tr} \sigma^{-2}Kb^{-1}fb^{-1}K^{\rm T}. \nonumber 
\end{eqnarray}
The second equality was obtained by noticing that the terms $j\ne k$
do not contribute on the average and that the average of $\Delta G_i^2$
is $\sigma_i^2$. Later we consider the average over $N_{\rm calc}$ MaxEnt 
calculations, each using data with the statistical accuracy $\sigma$. The 
statistical error $w_{\rm stat}$ is then reduced by a factor of $N_{\rm calc}$,
since $w_{\rm stat}$ refers to the square of the error in the output 
spectrum.  

For the systematic error we obtain
\begin{equation}\label{eq:a15}
w_{\rm syst}={\rm ln}({m\over A^{\rm exact}}) f \alpha  b^{-1}fb^{-1} 
\alpha f {\rm ln}({m\over A^{\rm exact}}).
\end{equation}
The results in Eqs.~({\ref{eq:a14}, \ref{eq:a15}) apply to the case 
when MaxEnt calculation is not iterated. For an iterative calculation
we use Eq.~({\ref{eq:e1}) below.

Fig.~\ref{fig:7} shows $w_{\rm stat}$ and $w_{\rm syst}$ as a function of 
$\alpha$ for different $\sigma$. The figure illustrates that $w_{\rm stat}$ behaves 
approximately as $1/\alpha$. This illustrates the importance of introducing 
entropy, i.e., using an $\alpha>0$. For $\alpha=0$, the matrix $b^{-1}$ is  
ill-behaved and the statistical error would be huge. Since $w_{\rm stat}$
depends only weakly on $\sigma$, it is not possible to make $w_{\rm stat}$
small for $\alpha=0$ by simply reducing $\sigma$ (within reasonable limits).
Introducing $\alpha>0$ regularizes $b$ and leads
to a manageable statistical error. The systematic error increases with 
$\alpha$ and there is therefore an optimal value of $\alpha$ where
the total error is minimum. The dependence of the systematic error 
on $\sigma$ is shown in Fig.~\ref{fig:8}. It behaves roughly as 
$\sqrt{\sigma}$. The optimal $\alpha$ therefore increases as $\sigma$ 
is reduced. It also increases as the default model is made more 
accurate, e.g., by replacing default model 1 by model 2.

\begin{figure}
{\rotatebox{-90}{\resizebox{6.cm}{!}{\includegraphics {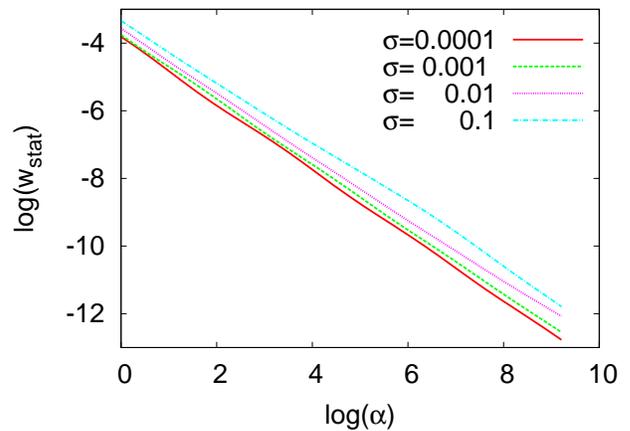}}}}
{\rotatebox{-90}{\resizebox{6.cm}{!}{\includegraphics {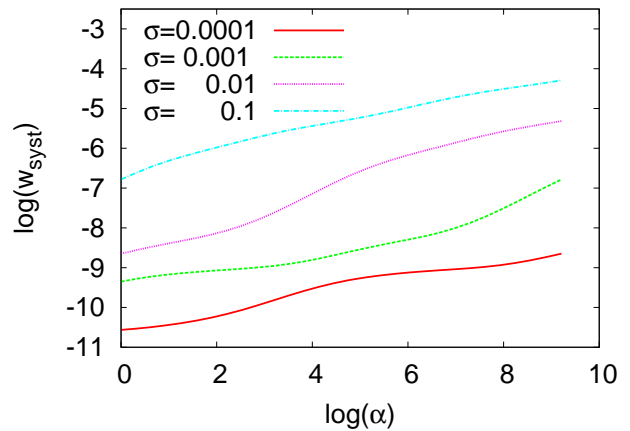}}}}
\caption{\label{fig:7}(color on-line) Statistical ($w_{\rm stat}$)    
[Eq.~(\ref{eq:a14})] and systematic ($w_{\rm syst}$) [Eq.~(\ref{eq:a15})]
errors for default model 1 as a function of $\alpha$ and for different 
values of $\sigma$. The parameters are $\beta=15$ and 
$N_{\tau}=60$.
}
\end{figure}

\begin{figure}
{\rotatebox{-90}{\resizebox{6.cm}{!}{\includegraphics {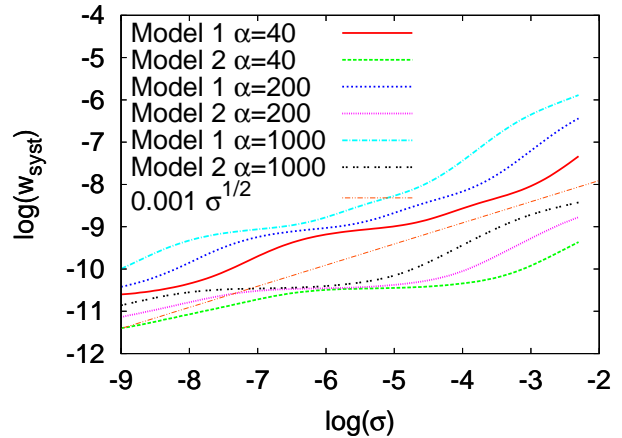}}}}
\caption{\label{fig:8}(color on-line) Systematic error ($w_{\rm syst}$)    
[Eq.~(\ref{eq:a15})] for default models 1 and 2  as a function of 
$\sigma$ and for different values of the $\alpha$. The straight 
line shows the curve $0.001 \sigma^{1/2}$, illustrating that 
$w_{\rm stat}$ is approximately proportional to $\sqrt{\sigma}$. 
The parameters 
are $\beta=15$ and $N_{\tau}=60$.
}
\end{figure}

Since $\omega=0$ is often of particular interest we use a logarithmic 
$\omega$-mesh. For the case  $0\le \omega \le \omega_{\rm max}$ we use
\begin{equation}\label{eq:a16}
\omega_i={\rm exp}[(i-1)dx+{\rm ln}\gamma]-\gamma,
\end{equation}
where $dx=[{\rm ln}(\omega_{\rm max}+\gamma)-{\rm ln}\gamma]/(N_{\omega}-1)$.
A small value of $\gamma$ leads to a smaller spacing of the points close to 
$\omega=0$. We have typically used $N_{\omega }=121$ points, $\gamma=0.5$, 
$\omega_{\rm max}=12$, $\beta=15$ and $N_{\tau}=60$. For simplicity, we 
assume that the statistical error is given by
\begin{equation}
\sigma_i=G_i \sigma,
\end{equation}
in terms of some overall accuracy $\sigma$. To perform these calculations
we have developed a MaxEnt code, which was found to give almost identical 
results to a code made available to us by Jarrell.\cite{Jarrell}

\section{Multiple MaxEnt calculations}\label{sec:3}

\begin{figure}
{\rotatebox{-90}{\resizebox{6.cm}{!}{\includegraphics {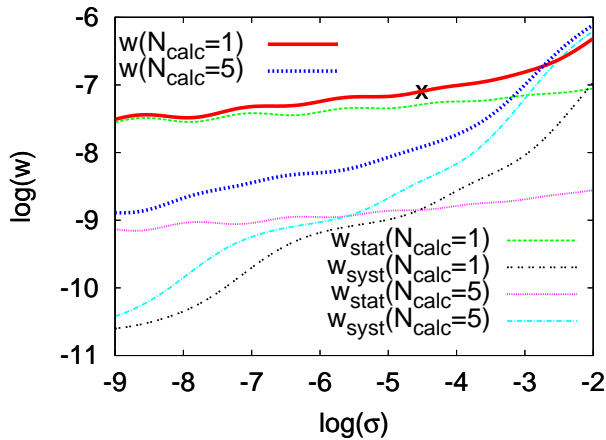}}}}
\caption{\label{fig:2}(color on-line) Statistical ($w_{\rm stat}$), 
systematic ($w_{\rm syst}$) and total ($w$) errors in a MaxEnt calculation 
for the spectrum in Fig.~\ref{fig:1}, default model 1, $\alpha=40$ and $\beta=15$ 
and 100 samples, each with the accuracy $\sigma$. The full thick (red) line
shows $w$ when one ($N_{\rm calc}=1$) MaxEnt calculation is performed 
for the average of over all data and the the thick broken (blue) line 
the result when $N_{\rm calc}=5$ MaxEnt calculations are averaged, each 
calculation using the average of $100/N_{\rm calc}$ samples. The cross 
corresponds to a historic MaxEnt calculation for $\sigma=0.01$, which 
gives $\alpha \approx 40$, used in the figure. The thick broken (blue) 
curve illustrates that a substantially lower error can be obtained by 
averaging 5 MaxEnt calculations. 
}
\end{figure}

A QMC calculation is arranged so that it gives a number of samples,
$\bar G_{\nu}(\tau_i )$, of $G(\tau_i)$. From these data one can calculate 
the statistical accuracy  $\sigma_i$, check if the data are Gaussian
and check for (undesirable) correlations between the noise for different
values of $\tau$.\cite{Jarrell} The data $\bar G_{\nu}(\tau_i)$ are then 
averaged over $\nu$ to obtain $\bar G(\tau_i)$ that is the input for 
the MaxEnt calculation, possibly after removing correlations between 
the noise at different $\tau$-points.\cite{Jarrell}

We now consider the case where we have 100 samples, each with the accuracy
$\sigma$. After averaging over all the samples, the accuracy of the
resulting data is $\sigma/\sqrt{100}=\sigma/10$. The value of $\alpha$ in
a classical MaxEnt calculation depends on the specific realization of the
noise. We therefore perform many calculations, each with a different 
realization of the noise,
and average over $\alpha$. For $\sigma=0.01$, $\beta=15$ and $N_{\tau}=60$, 
classical MaxEnt calculations using the spectrum in Fig.~\ref{fig:1} and 
the default model 1 then gave on the average  $\alpha \approx 40$.          

Fig.~\ref{fig:2} shows the statistical and systematic errors for a fixed
$\alpha=40$ as a function of $\sigma$ ($N_{\rm calc}=1$). The cross gives the 
total error of a classical MaxEnt calculation corresponding to 100 samples with 
$\sigma=0.01$. The figure illustrates that the statistical error is much larger 
than the systematic error for the MaxEnt calculation. This is also illustrated 
in Fig.~\ref{fig:3}a. This shows the result of 20 MaxEnt calculations with 
different realizations of the noise, each with 100 samples with the 
accuracy $\sigma$. The thick (red) line shows the exact spectrum. 
The calculated spectra (thin green lines) scatter strongly around the 
exact result, illustrating a large statistical error. On the average, 
these spectra also deviate somewhat from the exact result, the value of $\sigma(0)$ 
being slightly too small and the Hubbard peak being somewhat shifted towards 
lower energies, illustrating a small systematic error. 

\begin{figure}
{\rotatebox{-90}{\resizebox{5.3cm}{!}{\includegraphics {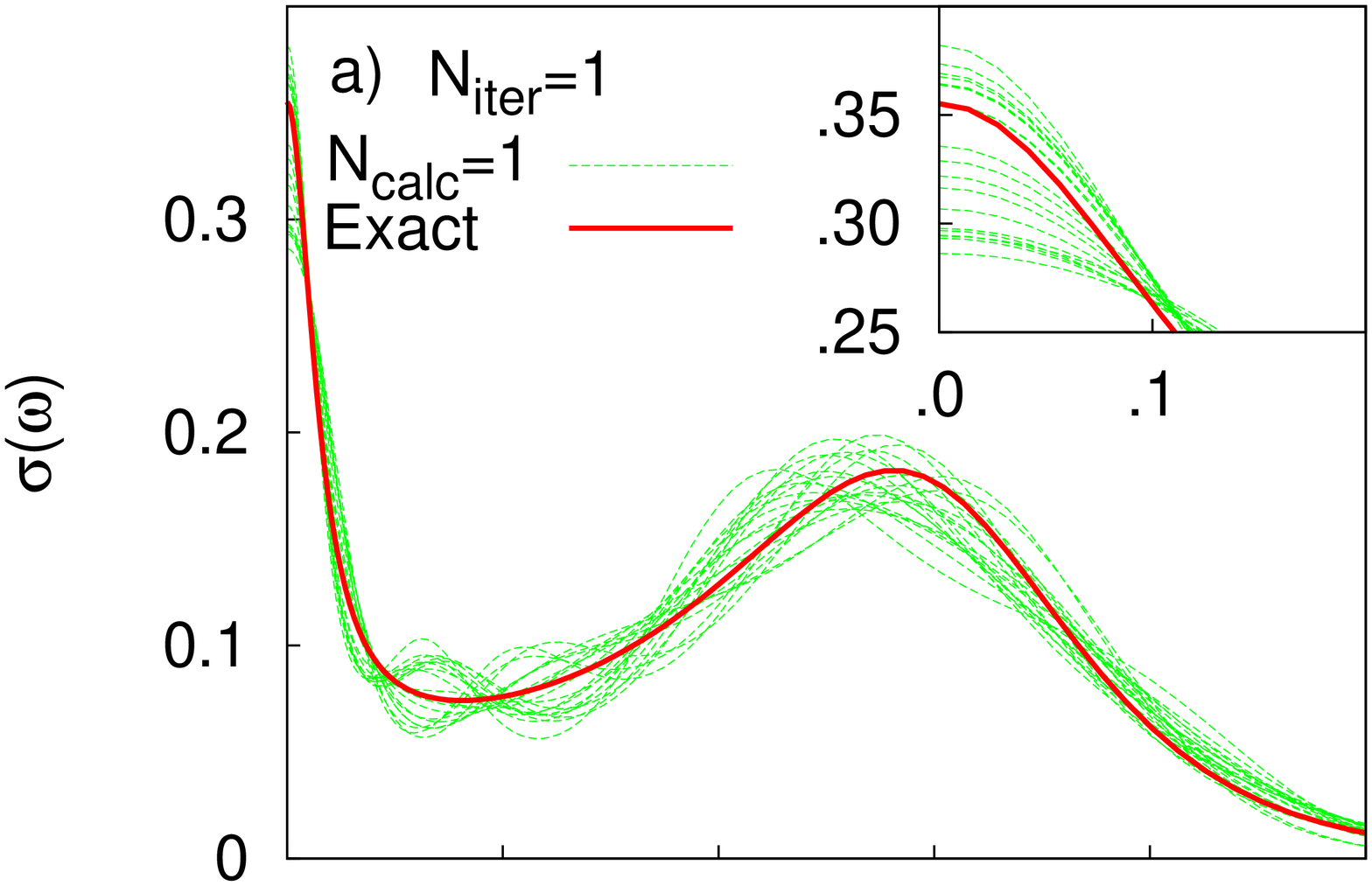}}}}
\vskip-1.15cm
{\rotatebox{-90}{\resizebox{5.3cm}{!}{\includegraphics {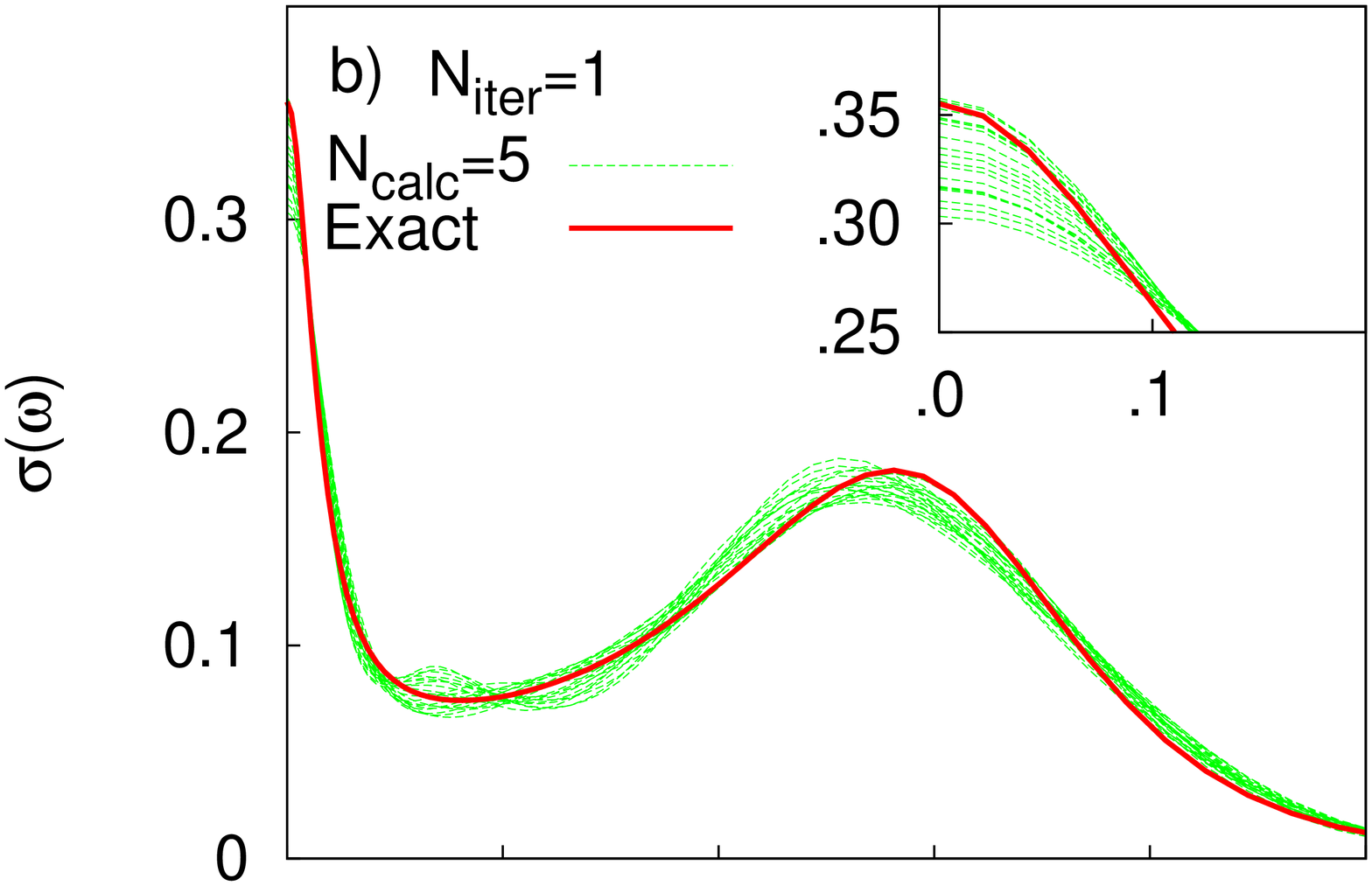}}}}
\vskip-1.15cm
{\rotatebox{-90}{\resizebox{5.3cm}{!}{\includegraphics {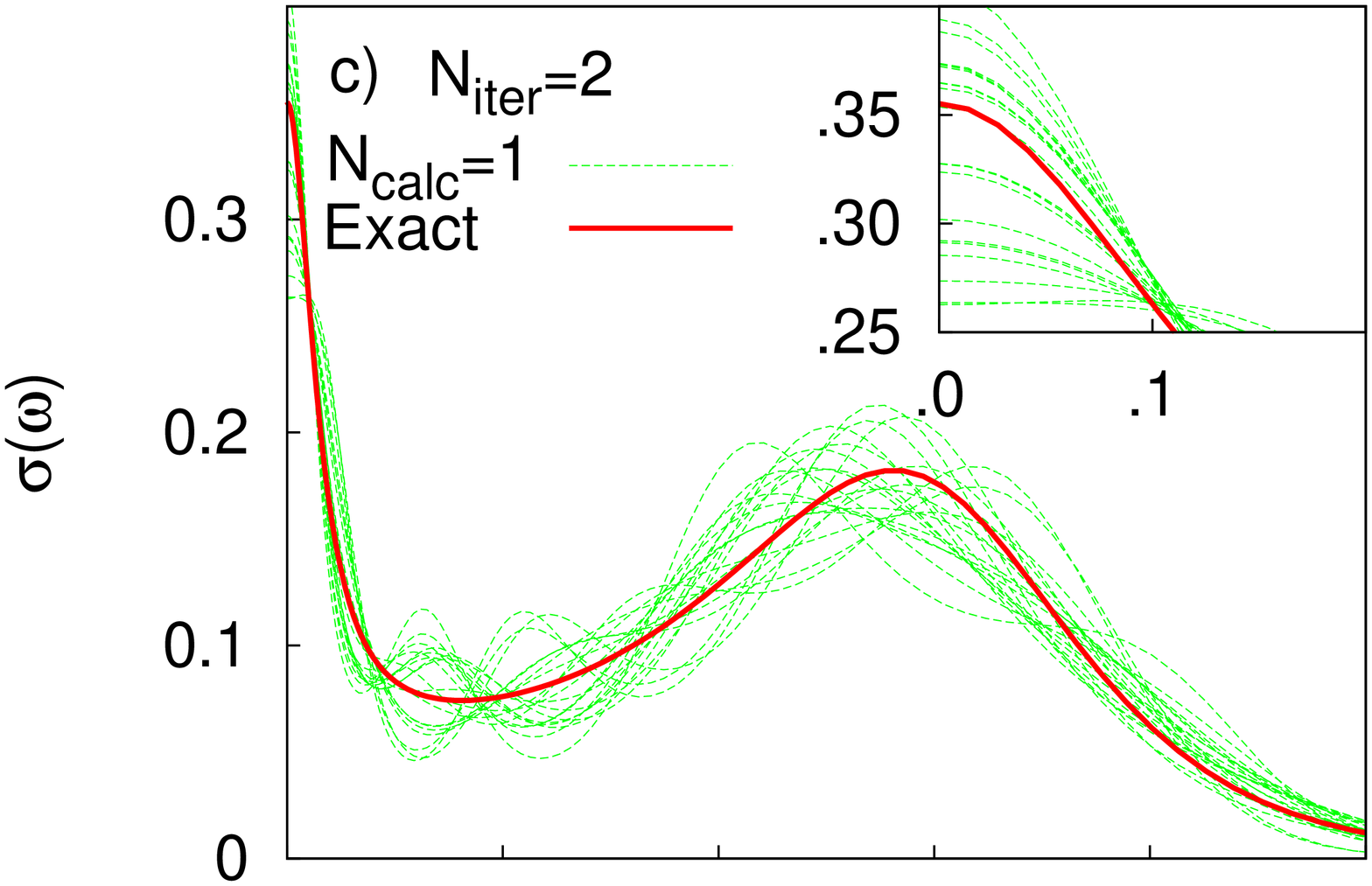}}}}
\vskip-0.90cm
{\rotatebox{-90}{\resizebox{5.3cm}{!}{\includegraphics {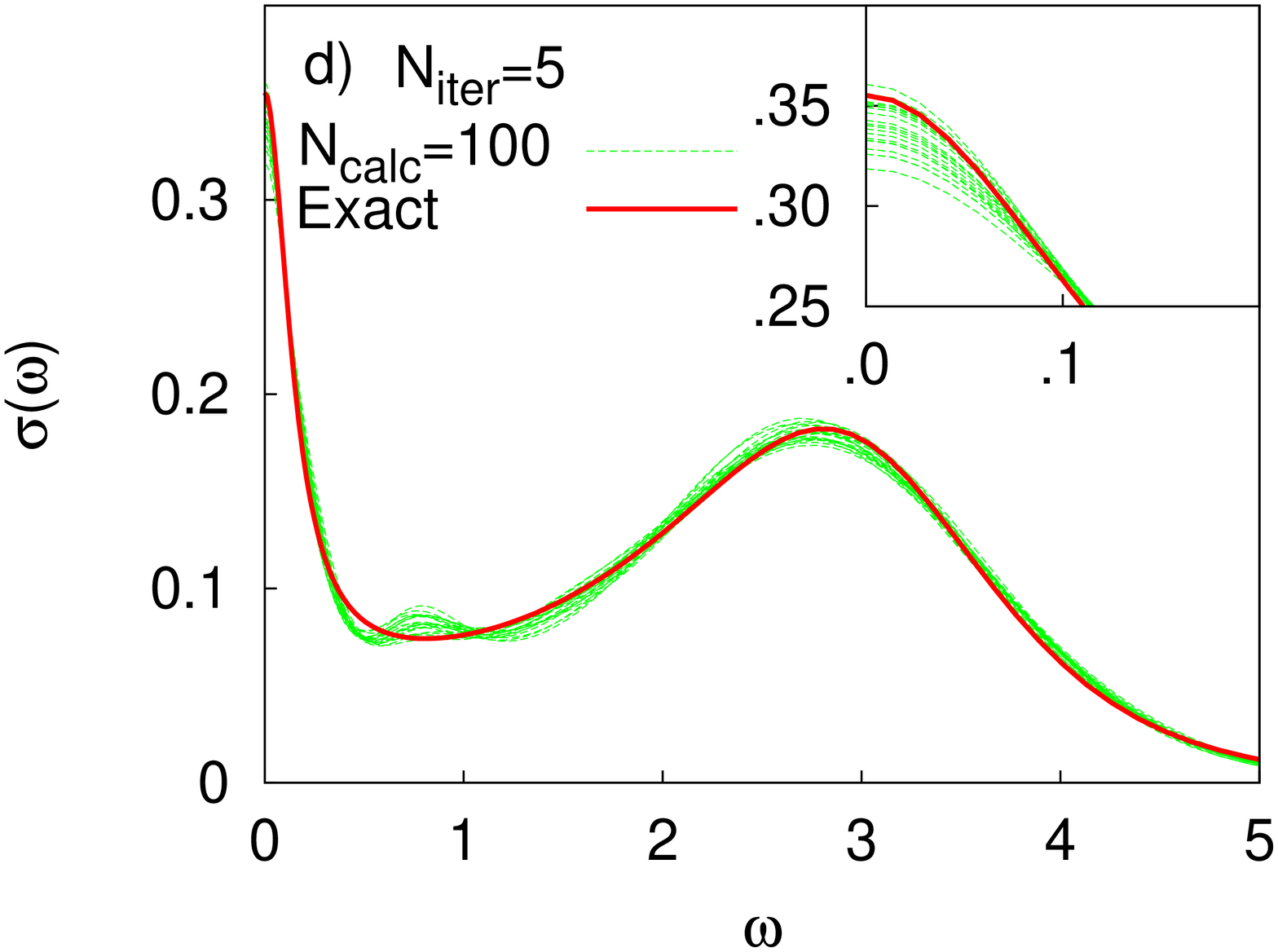}}}}
\caption{\label{fig:3}(color on-line) Optical conductivity calculated
for the default model 1 using different methods. 100 samples, each with
the accuracy $\sigma=0.01$ were given.  a) Each curve shows results of 
a classical MaxEnt calculation using an average of all 100 samples. The 
figure shows 20 such curves, each corresponding to a different realization 
of the noise. b) Each curve shows the average of $N_{\rm calc}=5$ MaxEnt
calculations using 100/$N_{\rm calc}$ samples. c) Each curve shows the results of 
iterating the calculations in a) once, using the output in a) as a default
function in the next MaxEnt calculation. d) Each curve shows the results of iterating 
MaxEnt calculations $N_{\rm iter}=5$ times. $N_{\rm calc}=100$ was used,
and the default function was obtained from the  average of these $N_{\rm calc}$ 
calculations. The parameters were $\beta=15$, $N_{\tau}=60$ and $\alpha=40$.
}
\end{figure}

\begin{figure}
{\rotatebox{-90}{\resizebox{5.3cm}{!}{\includegraphics {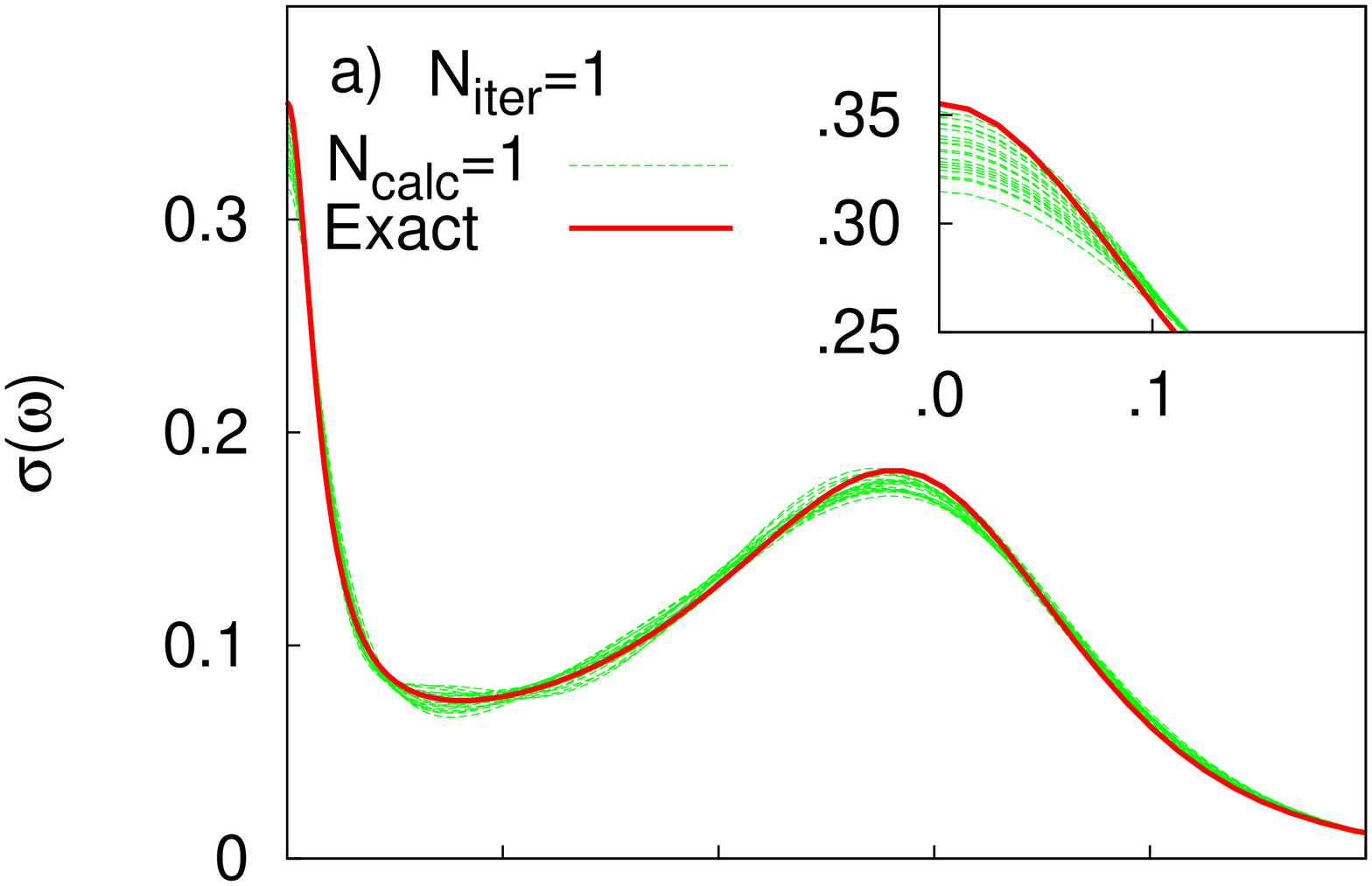}}}}
\vskip-1.15cm
{\rotatebox{-90}{\resizebox{5.3cm}{!}{\includegraphics {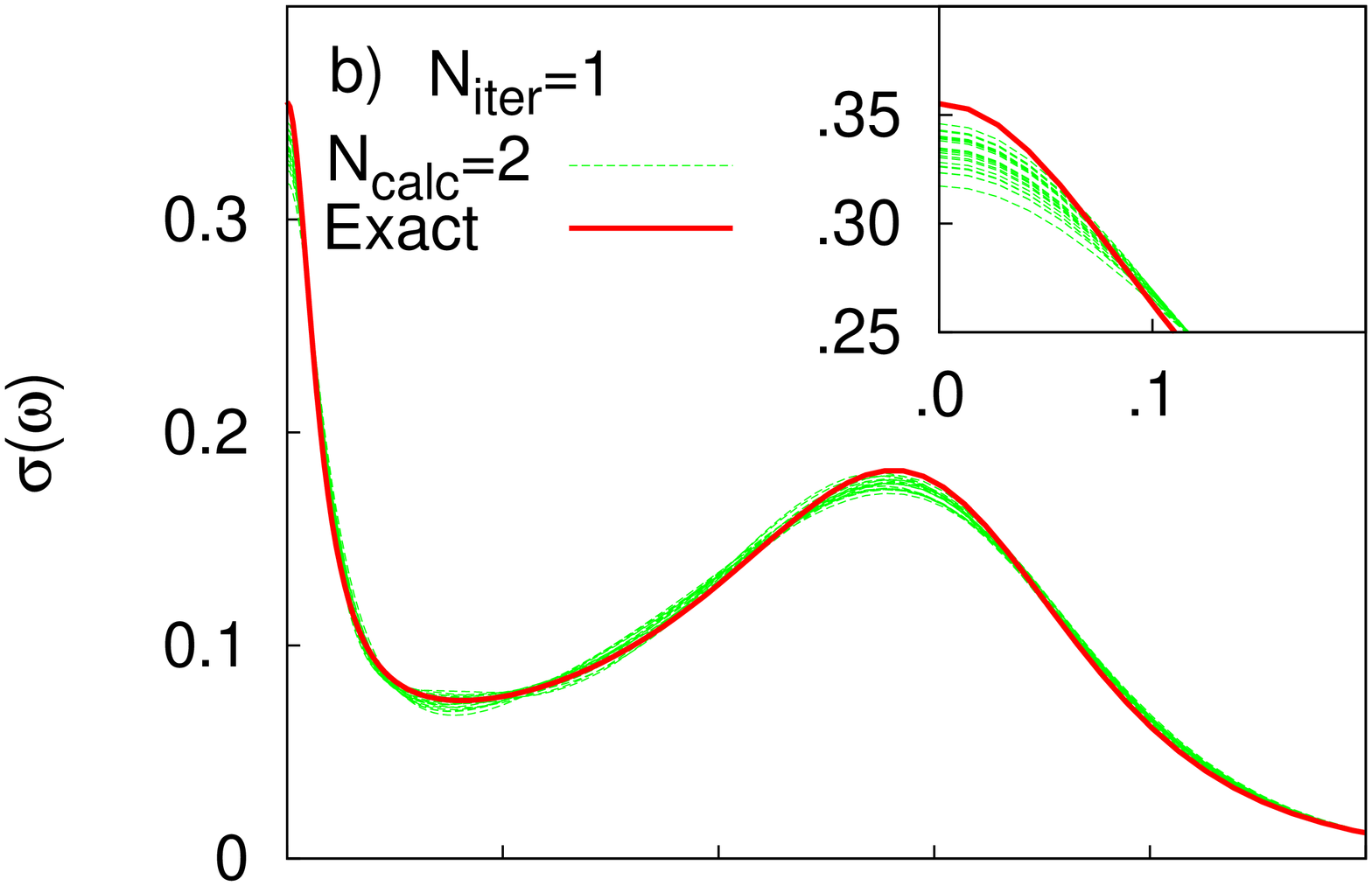}}}}
\vskip-1.15cm
{\rotatebox{-90}{\resizebox{5.3cm}{!}{\includegraphics {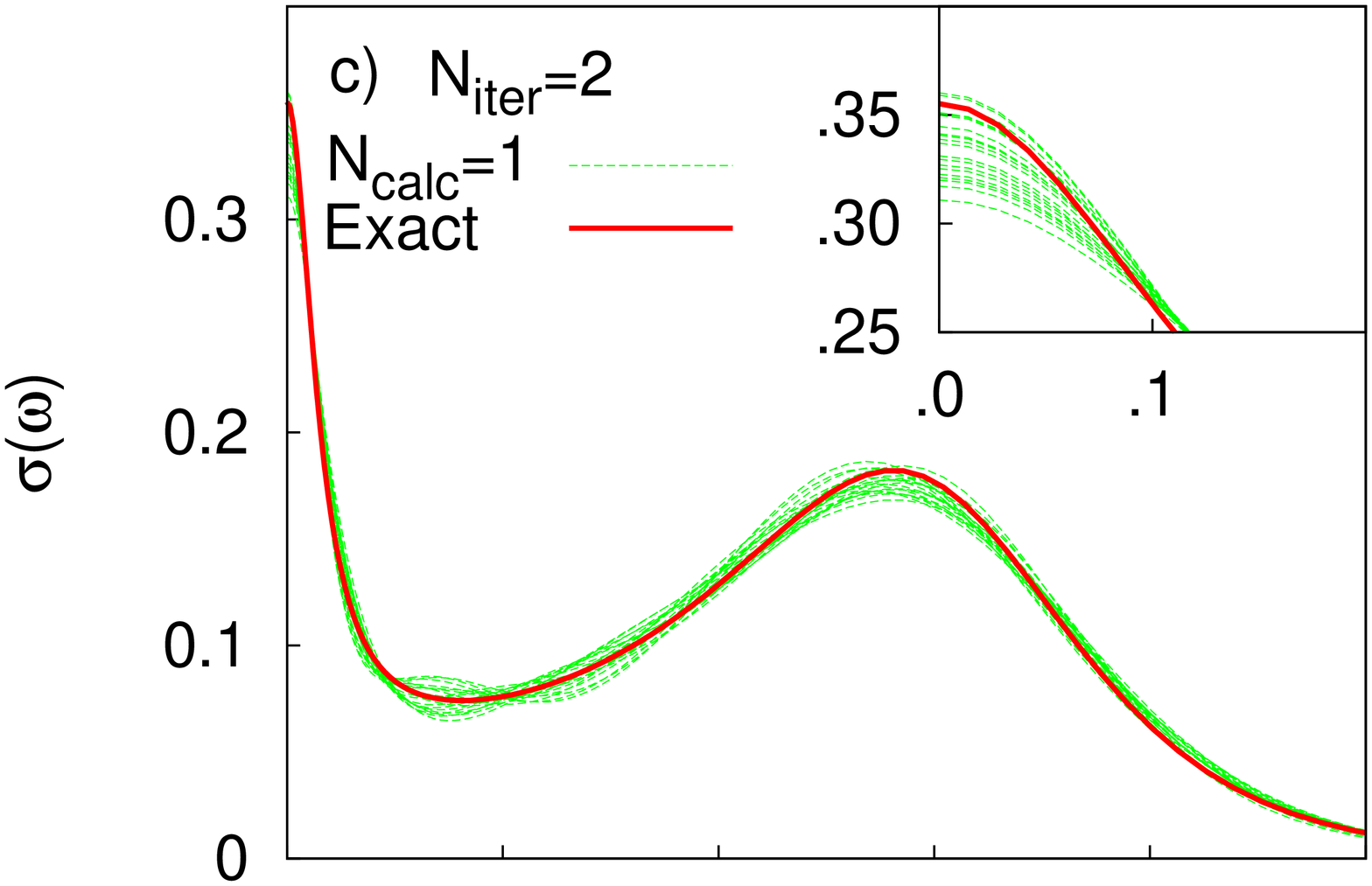}}}}
\vskip-0.90cm
{\rotatebox{-90}{\resizebox{5.3cm}{!}{\includegraphics {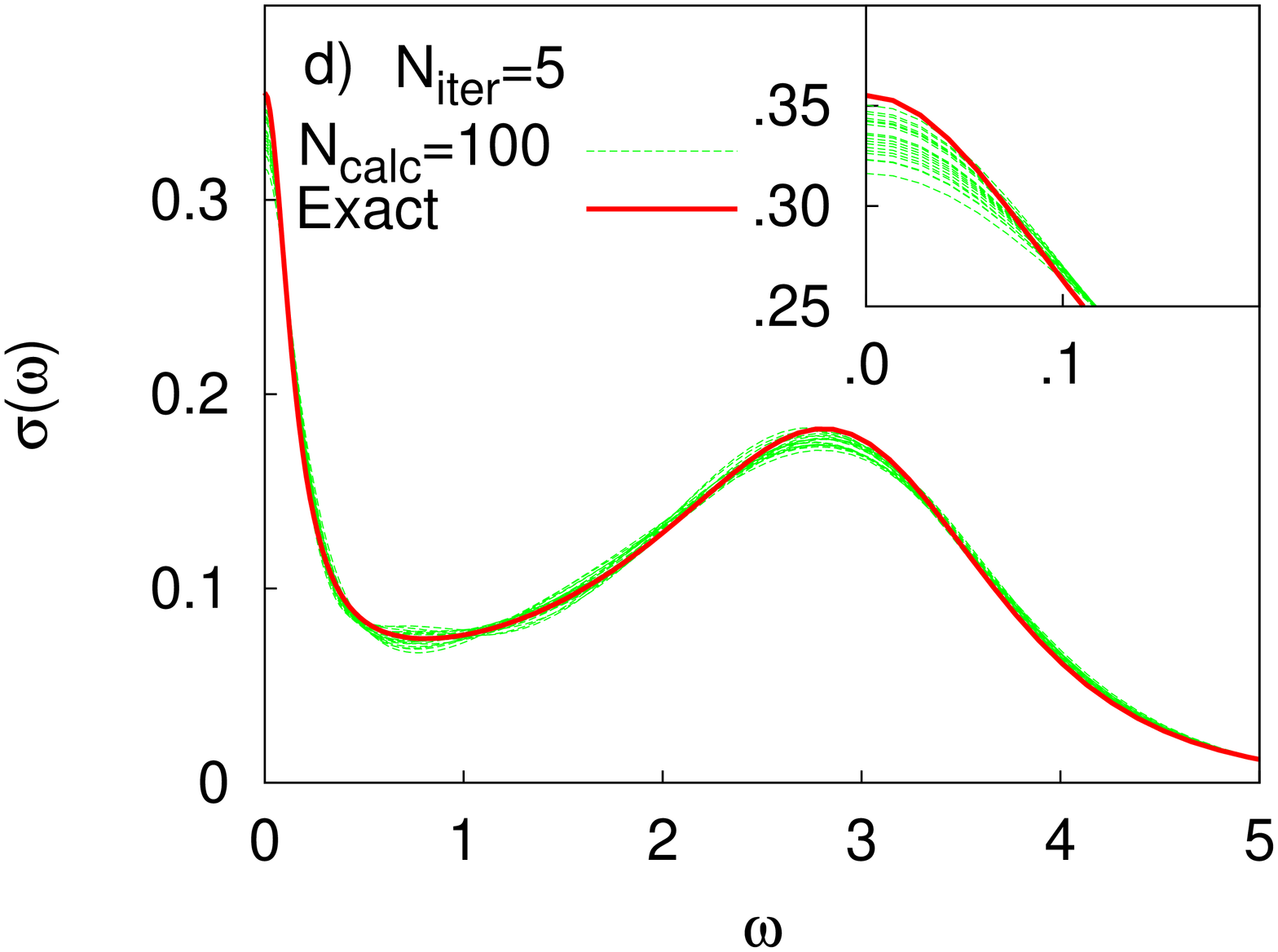}}}}
\caption{\label{fig:4}(color on-line) The same as Fig.~\ref{fig:3},
but starting from the default model 2 and using $\alpha=720$.  
}
\end{figure}

\begin{table}
\caption{\label{table:1}Statistical ($w_{\rm stat}$), systematic ($w_{\rm syst}$) and
total ($w$) error in MaxEnt calculations for the spectrum in Fig.~\ref{fig:1} and 
default models 1 or 2. 100 samples, each with the accuracy $\sigma$, were split
up in $N_{\rm calc}$ batches with 100/$N_{\rm calc}$ samples and used in 
$N_{\rm calc}$ calculations. The average of the output was used as a default model, performing 
$N_{\rm iter}$ iterations. The errors were obtain from Eqs.~(\ref{eq:a14}, \ref{eq:a15})
for $N_{\rm iter}=1$ and from Eq.~(\ref{eq:e1}) for $N_{\rm iter}>1$. The parameters 
were $\beta=15$ and $N_{\tau}=60$.
}
\begin{tabular}{ccccccc}
\hline
\hline
Default model & $\alpha$& $N_{\rm calc}$ & $N_{\rm iter}$ & $w_{\rm stat}$ & $w_{\rm syst}$ & $w$ \\ 
\hline
   1         &  40& 1        &   1  & 6.7 10$^{-4}$  & 1.4 10$^{-4}$  & 8.1 10$^{-4}$ \\
   1         &  40& 5        &   1  & 1.4 10$^{-4}$  &  2.1 10$^{-4}$  & 3.5 10$^{-4}$  \\
   1         &  40 & 1        &   2  & 14 10$^{-4}$  & 1.3 10$^{-4}$  & 15 10$^{-4}$  \\
   1         &  40& 100       &   4  & 0.4 10$^{-4}$     & 0.9 10$^{-4}$ & 1.3 10$^{-4}$ \\ 
   2         & 720&  1        &   1  & 4.2 10$^{-5}$ & 4.4 10$^{-5}$ &  8.6 10$^{-5}$ \\
   2         & 720&  2        &   1  & 2.3 10$^{-5}$ & 5.6 10$^{-5}$  & 7.9 10$^{-5}$  \\
   2         & 720&  1        &   2  & 9.7 10$^{-5}$ & 3.3 10$^{-5}$  & 13 10$^{-5}$ \\ 
   2         & 720&  20        &   8  & 2.4 10$^{-5}$& 3.9 10$^{-5}$ &  6.3 10$^{-5}$ \\
\hline
\end{tabular}
\end{table}

We next group the 100 samples in $N_{\rm calc}=5$ batches, each with 20 samples,
and perform $N_{\rm calc}$ MaxEnt calculations. The accuracy of the data in
these MaxEnt calculations is then only $\sqrt{N_{\rm calc}}\sigma/10$. This 
increases both the systematic and statistical errors somewhat. Averaging these 
calculation, however, reduces the statistical error by a factor $N_{\rm calc}$.
In Fig.~\ref{fig:2} this leads to a large net reduction in the statistical error, 
which more than compensates for the increase of the systematic error. This is
illustrated in Fig.~\ref{fig:3}b, which shows 20 such results, each one obtained 
by averaging $N_{\rm calc}=5$ MaxEnt calculations with $100/N_{\rm calc}$ samples,
but with different realizations of the noise.
The spread between the curves is substantially smaller ($w_{\rm stat}=0.00014$
vs. 0.00067) than in Fig.~\ref{fig:3}a, while the systematic error is somewhat
larger ($w_{\rm syst}=0.00021$ vs. 0.00014). This leads to a substantial 
improvement in the total error ($w=0.00035$ vs. 0.00081). These results are
also shown in Table~\ref{table:1}. 

The reason for this improvement is that that $w_{\rm stat}\gg w_{\rm syst}$
 in the MaxEnt calculation with $N_{\rm calc}=1$ and that $w_{\rm stat}$ 
and $w_{\rm syst}$ have different dependencies on $N_{\rm calc}$. 
For $\alpha=0$, Eq.~(\ref{eq:a14}) gives that $w_{\rm stat} \sim \sigma^2$. 
Splitting up the calculation in $N_{\rm calc}$ calculations makes the 
effective $\sigma$ a factor $\sqrt{N_{\rm calc}}$ larger, while averaging 
reduces the error by a factor $N_{\rm calc}$. The net result would be an 
unchanged statistical error. It is therefore crucial that the method has
been regularized by introducing an entropy. Fig.~\ref{fig:2} shows that
for realistic values of $\alpha$, $w_{\rm stat}$ actually has a quite weak 
dependence on $\sigma$, rather 
than behaving as $\sigma^2$. Splitting up the samples in several batches, 
and thereby reducing the accuracy of each batch, leads to a small increase 
in $w_{\rm stat}$ for each individual calculation. The averaging over 
$N_{\rm calc}$ calculations, however, reduces $w_{\rm stat}$ by a factor 
$N_{\rm calc}$. At the same time $w_{\rm syst}$ is increased, but only 
by approximately a factor $N_{\rm calc}^{1/4}$, since this quantity
behaves approximately as $\sqrt{\sigma}$ and $\sigma$ increases by a factor 
$\sqrt{N_{\rm calc}}$.

Fig.~\ref{fig:4} and Table~\ref{table:1} show the corresponding results
using default model 2. In this case the classical MaxEnt calculation 
chooses a value of $\alpha$ that makes $w_{\rm stat}$ and $w_{\rm syst}$
comparable. The gain from splitting up the samples and performing
several MaxEnt calculations is then much smaller.

We are now in the position to discuss the limits of accuracy that can
be obtained in this approach. We consider as before 100 samples with
the accuracy $\sigma=0.01$ and allow for any combination of $\alpha$,
$N_{\rm calc}\le 100$ and $N_{\rm iter}\le 40$. Starting from default model 1,
we obtain the results shown in Fig.~\ref{fig:6}a. The curve ``One calc.''
shows the result of a traditional MaxEnt calculation, using all the samples
in one calculation ($N_{\rm calc}=1$ and $N_{\rm iter}=1$). If a classical
MaxEnt calculation is performed, $\alpha \approx 20$ is obtained. This
result is shown by a cross. We can see that this value of $\alpha$ is not
optimal, and a larger $\alpha$ would have given a smaller error.
We next allow for $N_{\rm calc}>1$ calculations, each using $100/N_{\rm calc}$
samples. We find the value of $N_{\rm calc}$ which gives the best agreement
with the ``exact'' $\sigma(\omega)$. This (``Several opt.'') leads to a 
much higher accuracy for 
small values of $\alpha$. The curve is almost flat as a function of 
$\alpha$ over a substantial range. For large values of $\alpha$, 
$N_{\rm calc}=1$ gives the best accuracy, and the curve falls on top of 
the curve ``One calc.''.

\begin{figure}
{\rotatebox{-90}{\resizebox{6.cm}{!}{\includegraphics {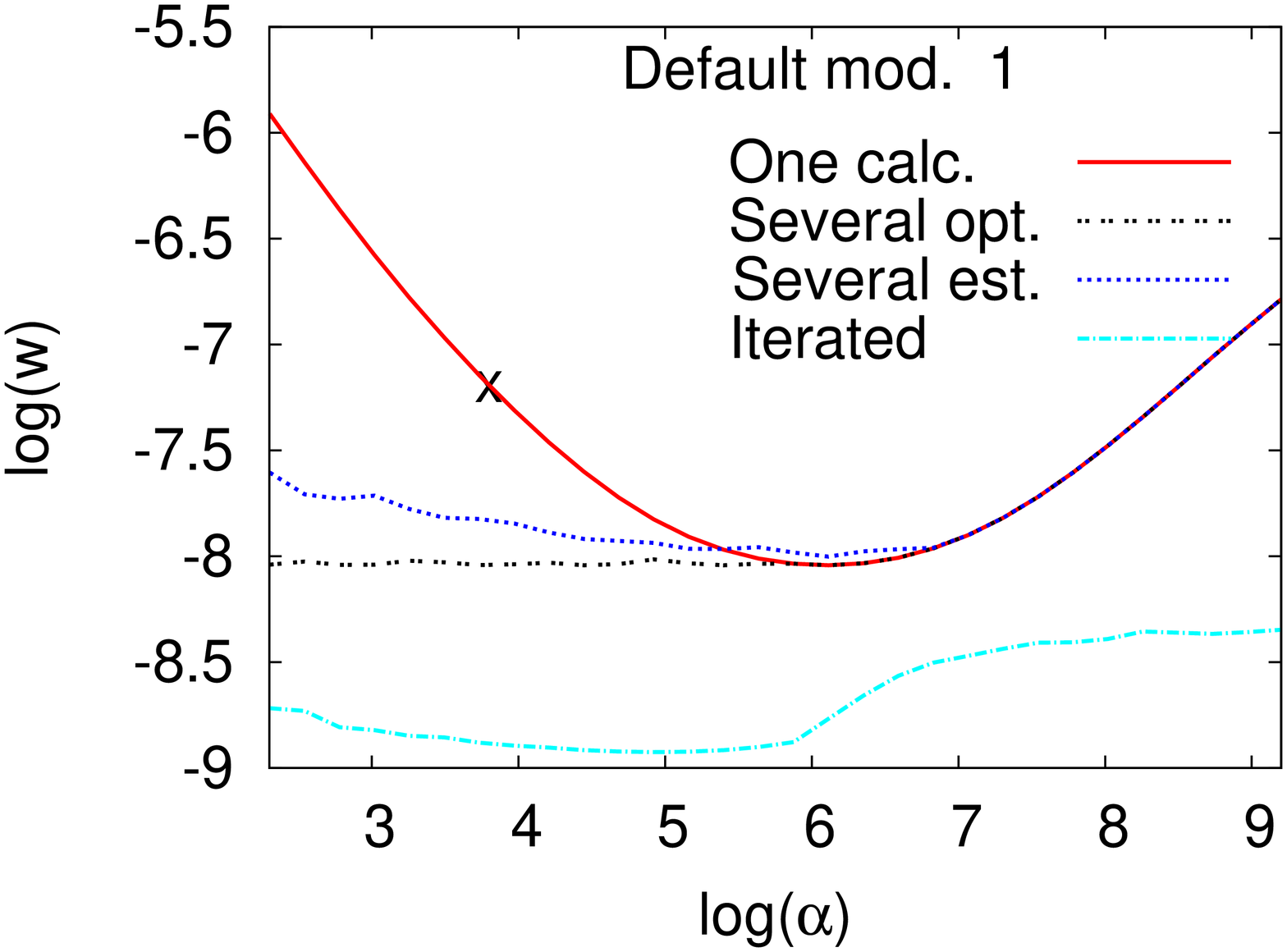}}}}
{\rotatebox{-90}{\resizebox{6.cm}{!}{\includegraphics {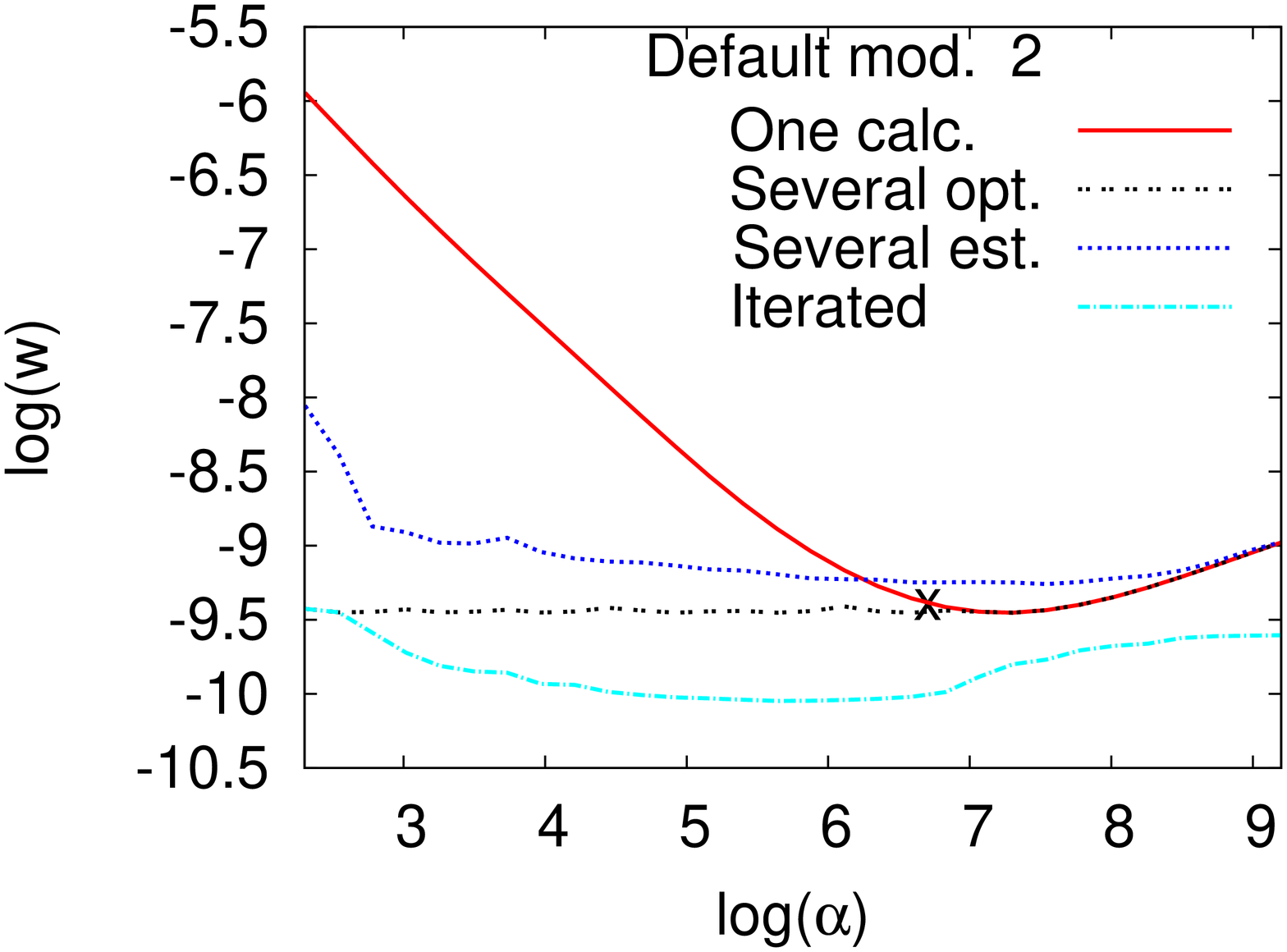}}}}
\caption{\label{fig:6}(color on-line)Accuracy $w$ of MaxEnt calculations 
for 100 samples, each with accuracy $\sigma$. "One calc." uses the average 
of all samples in one MaxEnt calculation. "Several opt." and ``Several est.''
split up the samples in several batches and averages the resulting MaxEnt 
calculations. "Several opt." does this in the optimal way and ``Several est.''
uses a prescription for finding the splitting when the exact result is not known.  
"Iterated" in addition uses the output spectral function as default model 
in an iterative approach. The cross shows the result of a classical MaxEnt 
calculation. a) shows results for default model 1 add b) for default model 2.
The parameters are $\beta=15$ and $N_{\tau}=60$. 
}
\end{figure}

To provide a criterion for how to split up the data in batches,
we consider the statistical error [(Eq.~\ref{eq:a14})] again.
As before we consider the case of $N_{\rm sample}$ samples, 
each with the accuracy $\sigma$, divided in $N_{\rm calc}$ batches 
with $N_{\rm sample}/N_{\rm calc}$ samples in each. We define the product 
\begin{equation}\label{eq:c1}
M(\sigma)=b^{-1}K^{\rm T}\sigma^{-2},
\end{equation}
where $b$ also depends on $\sigma$. The statistical error of the $k$th calculation 
is then written as
\begin{equation}\label{eq:c2}
\Delta A_i^{(k)}=\sum_{j=1}^{N_{\tau}}M_{ij}({\sigma_{N_{\rm calc}}})
{N_{\rm calc}\over N}\sum_{{\nu}=1}^{N/N_{\rm calc}}\Delta \bar G_j^{\nu+(k-1)N/N_{\rm calc}},
\end{equation}
where $\Delta \bar G^{\nu}$ is the error in the $\nu$th sample, and the statistical accuracy 
$\sigma_{N_{\rm calc}}=\sigma/\sqrt{N/N_{\rm calc}}$ enters due to the averaging over
$N/N_{\rm calc}$ samples. We then average over the $N_{\rm calc}$ calculations and 
obtain the error 
\begin{equation}\label{eq:c3}
\Delta A_i=\sum_{j=1}^{N_{\tau}}M_{ij}(\sigma_{N_{\rm calc}})
{1\over N}\sum_{{\nu}=1}^{N}\Delta \bar G_j^{\nu}.
\end{equation}
The average difference between two calculations with $N_{\rm calc}$ and 
$M_{\rm calc}$ batches can then be written as 
\begin{eqnarray}\label{eq:c4}
&&w_{MN}\equiv \sum_i[A_i(N_{\rm calc})-A_i(M_{\rm calc})]^2w_i=\\
&&\sum_{ij}w_i[{M_{ij}(\sigma_{N_{\rm calc}})\sigma_{N_{\rm calc}}\over \sqrt{N_{\rm calc}}}-
              {M_{ij}(\sigma_{M_{\rm calc}})\sigma_{M_{\rm calc}}\over \sqrt{M_{\rm calc}}}]^2. \nonumber
\end{eqnarray}
This result represents an average over many different realizations of 
the error $\Delta \bar G^{(k)}$. In addition to the statistical 
contribution to the difference there is a systematic contribution 
due to the error in the default function. We then compare calculations 
with $N_{\rm calc}=N_{\rm sample}$ and $M_{\rm calc}=N_{\rm sample}/2$ 
batches. In the second calculation the statistical error 
is larger and the systematic error is smaller. If the total difference
between the two calculations is larger than twice the expected statistical
difference, this suggests that the gain in the systematic error outweighs
the loss in the statistical error and the second calculation is accepted.
We then compare this calculation with a calculation with $N_{\rm sample}/4$ 
batches and if the latter is favorable the procedure is continued, 
considering $N_{\rm sample}/10$, $N_{\rm sample}/20$ and $N_{\rm sample}/50$ 
batches. The resulting accuracy is shown by the curves ``Several est.''
in Fig.~\ref{fig:6}. This curve is above the curve ``Several opt.'',
but the difference is not very large for most values of $\alpha$.

\section{Iterating MaxEnt}\label{sec:4}

Once a MaxEnt calculation has been performed, one can try to
improve the default function by using the output spectral function
as a new default function. Such an iterative approach, however, 
is usually not recommended. Fig.~\ref{fig:3}c shows the results 
of such calculations using default model 1. Indeed, the spread 
between between different calculations is larger than in the 
noniterated case in Fig.~\ref{fig:3}a, implying an increased statistical
error.

Eqs.~(\ref{eq:a14}, \ref{eq:a15}) used to calculate the statistical and
systematic errors for a noniterated default model are not appropriate 
in the case of iterations.  The reason is that the default model in 
this case contains statistical errors due to the iteration procedure. 
Instead we perform many calculations $N$ of the type shown in 
Fig.~\ref{fig:3}c, giving spectral functions $A_i^{\nu}$, $\nu=1, ..,N$. 
We then calculate
\begin{eqnarray}\label{eq:e1}
&&A^{\rm av}_i={1\over N}\sum_{\nu=1}^N A_i^{\nu} \nonumber \\
&&w_{\rm stat}={1\over N-1}\sum_{\nu=1}^N\sum_{i=1}^{N_{\omega}}(A_i^{\nu}-
A^{\rm av}_i)^2 f_i    \\
&&w_{\rm syst}=\sum_{i=1}^{N_{\omega}}(A^{\rm av}_i-A_i^{\rm exact})^2f_i
\nonumber
\end{eqnarray}
Due to nonlinearity, some of the statistical error actually shows up as
a systematic error in Eq.~(\ref{eq:e1}), but this is neglected in the following.

The results in Table~\ref{table:1}, shows that the statistical error is 
more than doubled after one iteration, while the systematic 
error is not correspondingly reduced. 

We next consider the case when the samples are split up in $N_{\rm calc}$ 
batches and default model 1 is used. Using $N_{\rm calc}=100$ and 
$N_{\rm iter}=4$ the total error is reduced, as is illustrated in 
Fig.~\ref{fig:3}d and Table~\ref{table:1}. By using $N_{\rm calc}=100$, 
we drastically reduce the statistical error. The following iterations 
increase the statistical error by a substantial factor, but it nevertheless 
remains small. At the same time the iterations reduce the systematic error, 
so that both are improved compared with the  noniterated case.
Fig.~\ref{fig:4}d shows similar results using default model 2. Since
this model is very close to the exact result, iterations now lead to
a small improvement, but even with such an accurate default function there 
is an improvement. In Fig.~\ref{fig:6}, the curve
``Iterated'' shows results when iteration is allowed ($N_{\rm iter} \le 40$).
This leads to a substantial improvement in the accuracy. Fig.~\ref{fig:6}b
shows similar results for default model 2.

\section{Quadratic entropy}\label{sec:5}

In Eq.~(\ref{eq:a5}) in Sec.~\ref{sec:2} we introduced the entropy, containing a
logarithm. As a result, the basic equations of MaxEnt are nonlinear, which makes
the analysis complicated. For this reason, we introduce a new definition of the 
entropy, which is used in this section for analyzing the behavior of MaxEnt. 
The expression in Eq.~(\ref{eq:a4}) is expanded to lowest (second) order in the 
deviation between the solution and the default function. We then {\it define} this
as the entropy, and use it in {\it this} section. This is then not an approximation
but simply a new method. This method has some problems. For instance,  it is
not guaranteed that the spectrum is positive. Therefore we do not recommend 
the use of this method for calculating spectra, but simply use it to analyze
MaxEnt.  

We define the entropy
\begin{equation}\label{eq:d1}
S=-{1\over 2}\sum_{i=1}^{N_{\omega}} f_i {(A_i^{(n)}-m_i^{(n)})^2 \over m_i^{(n)}},
\end{equation}
where we have allowed for the possibility of the MaxEnt calculation being iterated,
i.e., $m^{(n)}$ depends on the iteration $n$. The original default function
is $m^{(0)}$. This leads to the equations
\begin{eqnarray}\label{eq:d2}
&&a=K^{T}\sigma^{-2}\bar G +\alpha f \\ 
&&b^{(n)}=K^{T}\sigma^{-2}K +{\alpha f \over m^{(n)}} \nonumber                              
\end{eqnarray}
where matrix notations have been used and $\alpha f$ and $\alpha f / m^{(n)}$
are diagonal matrices. Then $A^{(n+1)}=[(b^{(n)}]^{-1}a$. 
The error in $A^{(n+1)}$ is 
\begin{equation}\label{eq:d4}
\Delta A^{(n+1)}=[b^{(n)}]^{-1}[K^T\sigma^{-2}\Delta \bar G 
+\alpha f {\Delta m^{(n)} \over m^{(n)}}],
\end{equation} where $\Delta m^{(n)}=m^{(n)}-A^{\rm exact}$.
We define
\begin{eqnarray}\label{eq:d5}
&&\delta G^{(n)}=\sqrt{m^{(n)} \over \alpha f}K^T\sigma^{-2}\Delta \bar G \nonumber \\
&&\delta A^{(n)}=\sqrt{\alpha f \over m^{(n)}} \Delta A^{(n)} \\
&&c^{(n)}=\sqrt{\alpha f \over m^{(n)}} [b^{(n)}]^{-1} \sqrt{\alpha f \over m^{(n)}}\nonumber
\end{eqnarray}
and a similar definition for $\delta m^{(n)}$ as for $\delta A^{(n)}$. $\sum_i[\delta A^{(n)}_i]^2$
contains the same integration factor $f_i$ as has been used earlier, but due to the factor
$1/m^{(n)}$ it gives more weight to errors where $m^{(n)}$ is small. We have 
\begin{equation}\label{eq:d6}
\delta A^{(n+1)}=c^{(n)}[\delta G^{(n)}+\delta m^{(n)}].
\end{equation}
We find the eigenvalues $\varepsilon_{\nu}^{(n)}$ and eigenvectors $|\nu^{(n)}\rangle$
of the symmetric matrix $c^{(n)}$. Introducing the expansion in these eigenvectors, 
$\delta A_{\nu}^{(n)}=\langle \nu^{(n)}|\delta A^{(n)}\rangle$ we obtain
\begin{equation}\label{eq:d7}
\delta A_{\nu}^{(n+1)}=\epsilon_{\nu}^{(n)}[\delta G_{\nu}^{(n)}+\delta m^{(n)}_{\nu}].
\end{equation}

The matrix $c$ can be rewritten as
\begin{equation}\label{eq:d8}
c^{(n)}=[\sqrt{m^{(n)} \over \alpha f}K^{T}\sigma^{-2}K\sqrt{m^{(n)} \over \alpha f}+1]^{-1}
\end{equation}
For the cases we have considered, the first matrix inside the bracket has a broad
range of positive eigenvalues, extending from eigenvalues much smaller than one
to much larger than one. As a result, the matrix $c^{(n)}$ is found to have some 
very small eigenvalues and many eigenvalues very close to one. This is illustrated
in Table~\ref{table:2}, which shows the lowest eigenvalues for the default model 1
and $\alpha=40$. 

Fig.~\ref{fig:5} shows the eigenfunctions $|\nu^{(0)}\rangle$ corresponding 
to the lowest eigenvalues in Table~\ref{table:1}. The lowest function is 
nodeless, and the higher functions have an increasing number of nodes. 
Functions with the eigenvalue very close to one oscillate so rapidly that 
the corresponding components of $\delta m^{(n)}_{\nu}$ and $\delta G_{\nu}^{(n)}$ 
tend to have small weights, as shown in Table~\ref{table:2}. As a comparison,
Table~\ref{table:2} also shows the expansion coefficients of the default
model 2 in the eigenfunctions obtained for the default model 1 and $\alpha=40$.

It is crucial for the success of a MaxEnt calculation that the coefficients
$\delta m^{(n)}_{\nu}$ and $\delta G_{\nu}^{(n)}$ are typically small for
$\varepsilon_{\nu}^{(n)}$ close to one. From 
Eq.~(\ref{eq:d7}) it follows that errors $\delta m^{(n)}_{\nu}$ and 
$\delta G_{\nu}^{(n)}$ corresponding to eigenvalues $\varepsilon_{\nu}$  
much smaller than one give a strongly reduced contribution to the error 
$\delta A_{\nu}^{(n+1)}$, while errors   corresponding to the eigenvalue 
one are not reduced at all. For these components the deviation of the 
default model from the true result are taken over completely. 

At the same time this sets the limits for MaxEnt calculations. A MaxEnt calculation 
fails if $A(\omega)$ has structures on such a small energy scale that there 
are important expansion coefficients $\delta A_{\nu}^{(n+1)}$ corresponding 
to eigenvalues close to one, since the  MaxEnt calculation gives no additional 
information about these components. This also shows the danger of putting in 
too much structure on a small energy scale in the default function. This would 
make components $\delta m^{(n)}_{\nu}$ corresponding to $\varepsilon_{\nu} \approx 1$
important and the MaxEnt calculation would not remove them from $A_{\nu}^{(n+1)}$, even
if there is no support for such components in the data. 

The results in Figs.~\ref{fig:3} and \ref{fig:4} show a beating pattern,
where the different calculations agree approximately for certain values of $\omega$.
This must be related to the noise in the input data, since this is what differs 
between the calculations. The reason can be seen from Table~\ref{table:2} and 
Fig.~\ref{fig:5}. The contribution of the noise to the output is given by 
$\varepsilon_{\nu} \delta G_{\nu}^{(n)}$. This contribution comes mainly from 
the eighth and ninth eigenvalues. The corresponding eigenfunctions in Fig.~\ref{fig:5}
have their zeros approximately where the deviations between the calculations 
in Fig.~\ref{fig:3} are small, although the agreement is not perfect. The
reason is probably the nonlinearity due to the logarithm in Eq.~(\ref{eq:a8}).
For instance, if the logarithm is expanded to second order, the resulting product
of two functions generates functions with more nodes than either of the two functions. 
As a result we find that $\delta A_{\nu}^{(n+1)}$ has appreciable errors also
for components with a few more nodes than the eighth and ninth eigenfunctions.
This then shifts the beating pattern slightly towards lower energies.

We introduce the projection operator
\begin{equation}\label{eq:d7a}
P^{(n)}=\sum_{\nu} |\nu^{(n)}\rangle \langle \nu^{(n)} | 
\Theta(\varepsilon_0-\varepsilon_{\nu}^{(n)}),
\end{equation}
where the $\Theta$-function selects states with eigenvalues smaller than 
$\varepsilon_0<1$. Eq.~(\ref{eq:d7}) can now be iterated. If we assume that 
$\epsilon_{\nu}^{(n)}$ is independent of $n$, which is a good approximation, 
we obtain
\begin{equation}\label{eq:d9}
\delta A_{\nu}^{(n+1)}=\left \{\begin{array}{cc}\sum_{i=1}^{n+1}\varepsilon_{\nu}^i
\delta G_{\nu}^{(0)}+ \varepsilon_{\nu}^{n+1} \delta m^{(0)}_{\nu}
& {\rm for} \ \varepsilon_{\nu} \le \varepsilon_0 \\
\varepsilon_{\nu}[\delta G_{\nu}^{(0)} + \delta m^{(0)}_{\nu}] & {\rm for} \ 
\varepsilon_{\nu} >   \varepsilon_0
\end{array}\right.
\end{equation}
This illustrates how iteration reduces the systematic error for components 
with $\varepsilon_{\nu} \le \varepsilon_0$, but increases the statistical 
error. Whether iteration pays off then depends on the relative size of the 
statistical and systematic errors and the choice of $\varepsilon_0$.
In this linearized version, however, it does not pay off to include 
all states in the projection operator (leading to $P^{(n)}\equiv 1$).

For the nonlinear case, the behavior is a bit different. From the expression
for the error in Eq.~(\ref{eq:a8}), it follows that ln ($m/A^{\rm exact})$
enters. Expanding the logarithm leads to terms with products of eigenfunctions
of the type in Fig.~\ref{fig:5}. Such products couple to higher eigenfunctions 
with more nodes. The result is that the error of a certain $\nu$-component 
of ln($m/A^{\rm exact})$ depends not only on the error of that $\nu$-component
of $m$ but also on the errors of other components, in particular lower ones.  
Whether the errors from the different contributions add constructively or 
destructively depends on the specifics of the model. For the cases we considered
the contributions to the higher components often add destructively. Then 
it can be more favorable to iterate all components rather than just the ones that  
would be favorable according to Eq.~(\ref{eq:d9}). For the cases we 
have studied, this has usually been the case and this is the approach 
we used in Sec.~\ref{sec:4}.

\begin{table*}
\caption{\label{table:2}Lowest eigenvalues $\varepsilon_{\nu}$ of the matrix $c^{(0)}$ 
[Eq.~(\ref{eq:d5})] and the corresponding amplitudes $\delta m^{(0)}_{\nu}$ (1) and 
$\delta G^{(0)}_{\nu}$ for default model 1. The values of $\delta m^{(0)}_{\nu}$ (2) 
for default model 2 (expanded in the functions corresponding to default model 1) and 
the expansion coefficients of $\sigma(\omega)$ are also shown. The larger eigenvalues 
are all close to unity, the amplitudes of the corresponding $\delta G^{(0)}_{\nu}$ are 
all smaller than 2 10$^{-4}$.                  
The corresponding values of $\delta m^{(0)}_{\nu}$ are also fairly small, smaller than 0.1
for default model 1 and smaller than 0.02 for default model 2.
We used $\beta=15$, $N_{\omega}=121$,  $\omega_{\rm max}=12$, $\alpha=40$ and $\sigma=0.001$.
}
\begin{tabular}{cccccccccccccc}
\hline
\hline
$\varepsilon_{\nu}$ & .3 10$^{-6}$ & .2 10$^{-5}$ & .6 10$^{-5}$ & .30 10$^{-4}$ & .2 10$^{-3}$ &
.002 & .016 & .166 & 0.747 & 0.982 & 0.999 &1.000 & 1.000\\
 $\delta m^{(0)}_{\nu}$ (1) &-.92  & -.81 & -.28 & -1.7 &  1.1 & -.01 & -1.1 & -.17 & .31 & -.08 & -.33 & -.15 & .007 \\ 
 $\delta m^{(0)}_{\nu}$ (2) & .80  & -.32 &  .06 & -.08 &  .08 &  .03 & -.19 & -.11 & .07 &  .03 & -.10 & -.08 &-.007 \\ 
$\sigma_{\nu}$  &  3.1 & -2.4 & 2.9 & 1.5 & -1.3 & .09 & 1.0 & .18 & -.31 & .08 & .33 & .16 & -.007 \\
$\sqrt{\langle (\delta G^{(0)}_{\nu})^2 \rangle}$ & 1.6 10$^{3}$ &675    &371  & 176  & 65   &
23  & 7.90& 2.1 & .55 & 0.14 & .03 & .006 & .001\\
$\varepsilon_{\nu} \sqrt{\langle (\delta G^{(0)}_{\nu})^2 \rangle}$ & .6 10$^{-3}$ & .002 & .002 & .005&
 .014 & .038 & .128 & .349 & .411 & .134 & .030 & .006 & .001 \\

\hline
\end{tabular}
\end{table*}

\begin{figure}
{\rotatebox{-90}{\resizebox{6.cm}{!}{\includegraphics {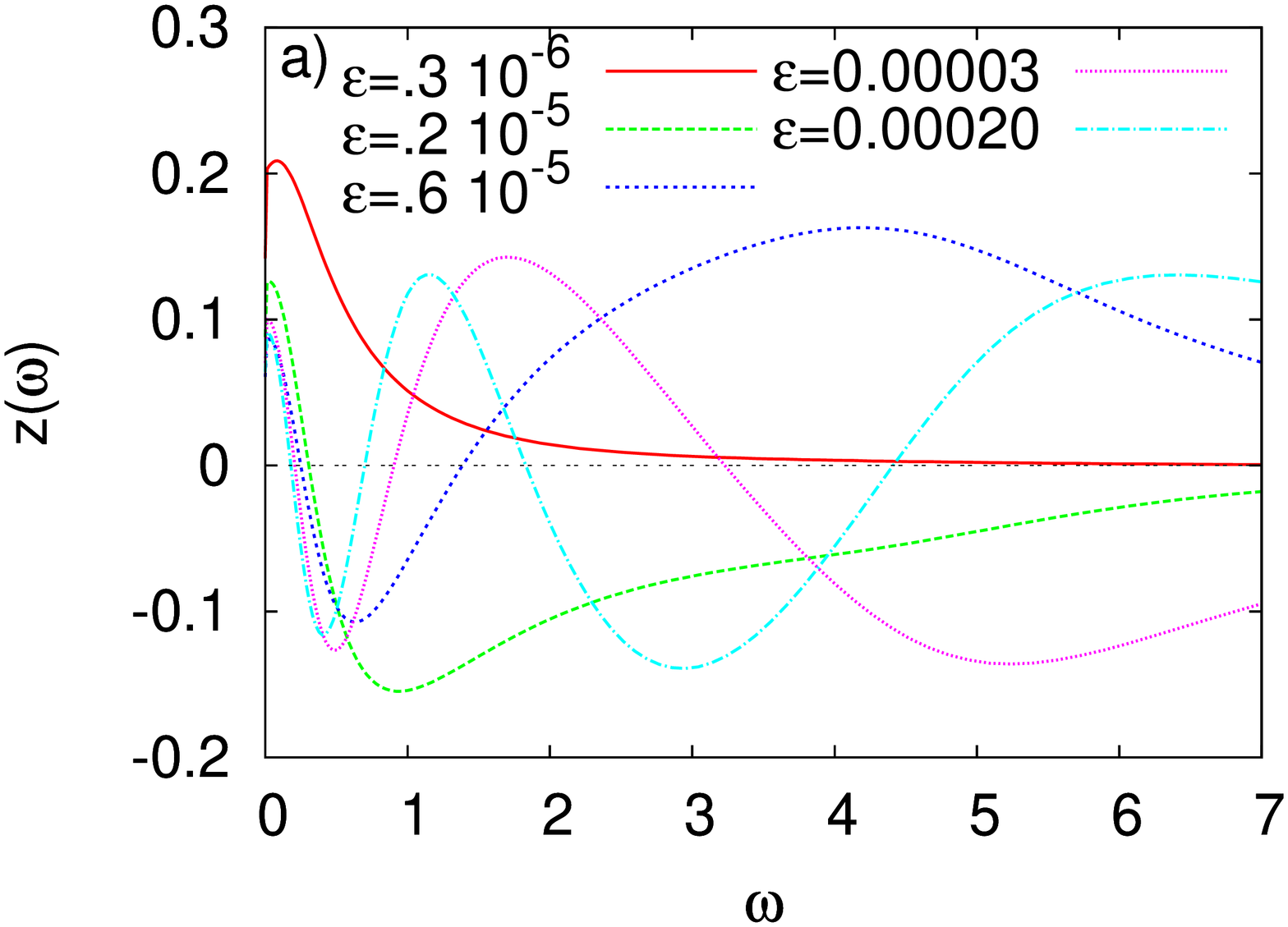}}}}
{\rotatebox{-90}{\resizebox{6.cm}{!}{\includegraphics {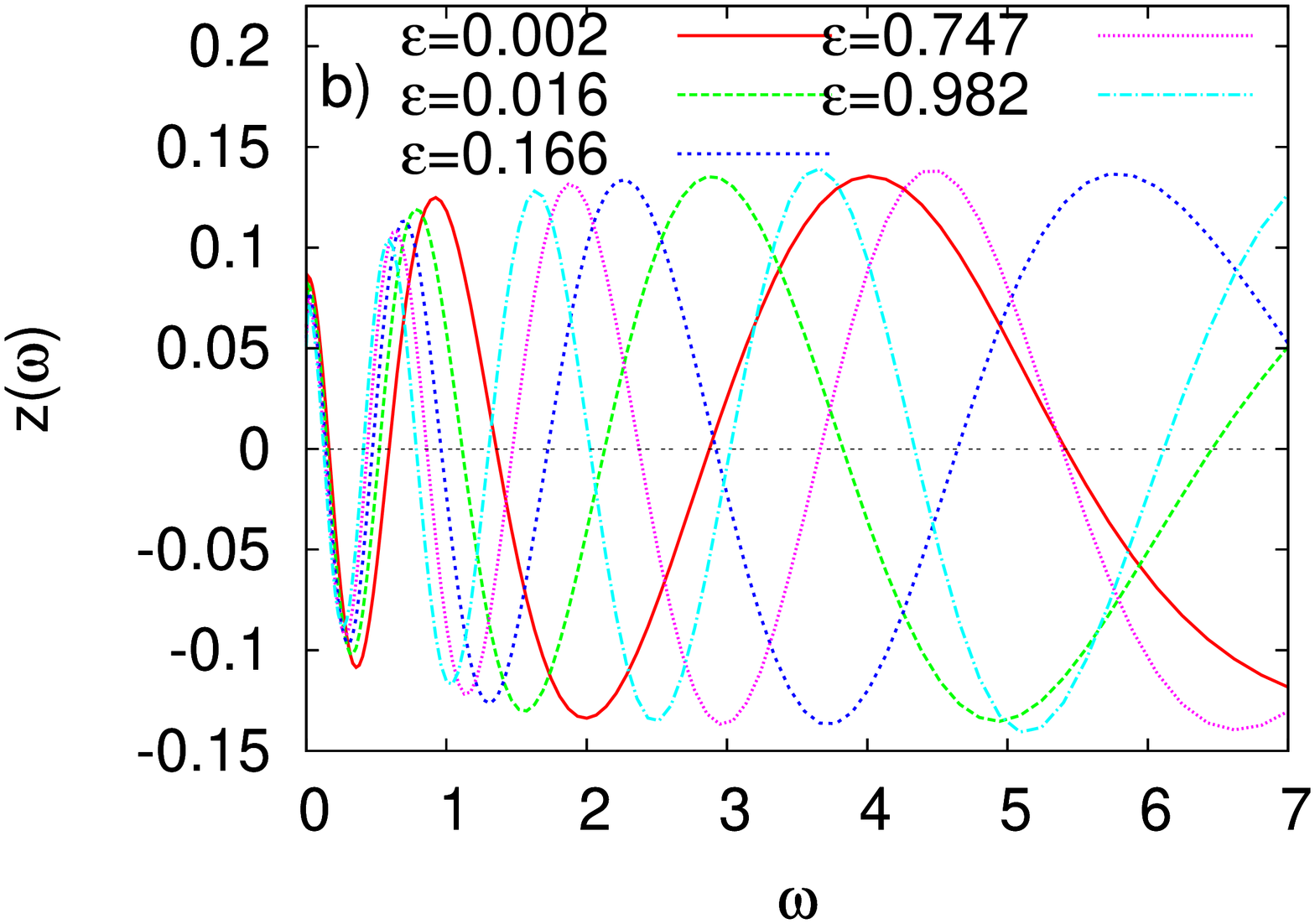}}}}
\caption{\label{fig:5}(color on-line) Eigenfunctions to the matrix $c^{(n)}$
in Eq.~(\ref{eq:d5}) corresponding to the 10 lowest eigenvalues. The figure
illustrates how eigenfunctions corresponding to an eigenvalue close to one
oscillate very rapidly. 
}
\end{figure}

\section{Summary}\label{sec:6}
We have analyzed the MaxEnt approach for analytical continuation, defining    
a statistical error, due to noise in the imaginary axis input data, and a systematic
error, due to errors in the default function entering the entropy. The classical 
method for choosing the weight $\alpha$ of the entropy can lead to a nonoptimal 
choice, reducing the systematic error at the cost of making the statistical error
unnecessarily large. We find that the statistical error
can be reduced by splitting up the data in batches. A MaxEnt calculation is performed 
for each batch and the result is averaged. This approach increases the systematic
error but the total error can be reduced. We have also studied an iterative approach,
where the output spectrum is used as default function in a new MaxEnt calculation.
We find that a straightforward application of this approach often gives worse results
due to a rapid increase of the statistical error. By splitting up the data in batches,
the statistical error can be reduced sufficiently that this is less serious. The reduction
of the systematic error can then outweigh the increase of the statistical error.

To analyze MaxEnt method, we have studied a linearized version of the problem. In this   
formalism it is easier to see how the statistical error propagates, in particular in the
case of iterations. One can also see how certain deviations of the default function
from the exact result have little influence on the output, while others fully show 
up in the output. This illustrates the danger of having a default function
with too much structure.

While this paper shows the potential for improving the MaxEnt method, it is harder 
to provide prescriptions for how to use this. In Sec.~\ref{sec:3} we provided a 
prescription for how to split the data in batches, which we have found to 
often work fairly well for a give value of $\alpha$. This method makes the resulting
error less sensitive to the optimization of alpha.  Alternatively, one can 
simply split the data in, say 10, batches. For each batch the classical method
of determining $\alpha$ is used and the resulting MaxEnt results are averaged.
This approach typically improves the accuracy of the output spectrum. In particular,
it reduces the risk of finding spurious structures due to
overfitting of noisy data, while some real structures can be lost in this approach. 

\section{Acknowledgments}
We would like to thank M. Jarrell for making his MaxEnt program available.
One of us (GS) wants to thank for support through the FWF ``Lise-Meitner'' 
grant n. M1136.

\end{document}